\documentclass[conference]{IEEEtran}
\IEEEoverridecommandlockouts
\usepackage{cite}
\usepackage{amsmath,amssymb,amsfonts}
\usepackage{algorithmic}
\usepackage{algorithm}
\usepackage{graphicx}
\usepackage{textcomp}
\usepackage{xcolor}
\graphicspath{ {./figures/} }
\def\BibTeX{{\rm B\kern-.05em{\sc i\kern-.025em b}\kern-.08em
    T\kern-.1667em\lower.7ex\hbox{E}\kern-.125emX}}
\begin{document}

\title{Adaptive Generalized Logit-Normal Distributions for Wind Power Short-Term Forecasting
\thanks{The research leading to this work is being carried out as a part of the Smart4RES project (European Union’s Horizon 2020, No. 864337). The
sole responsibility of this publication lies with the authors. The European
Union is not responsible for any use that may be made of the information
contained therein. The authors additionally acknowledge {\O}rsted for providing the data for the Anholt offshore wind farm.}
}

\author{\IEEEauthorblockN{Amandine Pierrot}
\IEEEauthorblockA{\textit{Technical University of Denmark}\\
Kgs Lyngby, Denmark \\
amapi@dtu.dk}
\and
\IEEEauthorblockN{Pierre Pinson}
\IEEEauthorblockA{\textit{Technical University of Denmark}\\
Kgs Lyngby, Denmark \\
ppin@dtu.dk}
}

\IEEEoverridecommandlockouts
\IEEEpubid{\makebox[\columnwidth]{978-1-6654-3597-0/21/\$31.00~\copyright2021  European Union \hfill} \hspace{\columnsep}\makebox[\columnwidth]{ }}

\maketitle

\IEEEpubidadjcol

\begin{abstract}
There is increasing interest in very short-term and higher-resolution wind power forecasting (from mins to hours ahead), especially offshore. Statistical methods are of utmost relevance, since weather forecasts cannot be informative for those lead times. Those approaches ought to account for the fact wind power generation as a stochastic process is non-stationary, double-bounded (by zero and the nominal power of the turbine) and non-linear. Accommodating those aspects may lead to improving both point and probabilistic forecasts. We propose to focus on generalized logit-normal distributions, which are naturally suitable and flexible for double-bounded and non-linear processes. Relevant parameters are estimated via maximum likelihood inference. Both batch and online versions of the estimation approach are described -- the online version permitting to additionally handle non-stationarity through parameter variation. The approach is applied and analysed on the test case of the Anholt offshore wind farm in Denmark, with emphasis placed on 10-min-ahead forecasting.

\end{abstract}

\begin{IEEEkeywords}
Wind power, Probabilistic forecasting, Dynamic models, Bounded time-series
\end{IEEEkeywords}

\section{Introduction}
Forecasting is of utmost importance to the integration of renewable energy into power systems and electricity markets. The attention of energy forecasting has increased tremendously over the years \cite{Hong2020}. For instance, thinking of short-term operational problems, transmission system operators (TSOs) have to operate reserves optimally to keep the system in balance at reasonable costs. Indeed, in Denmark, the TSO has some time argued the 10-min lead time as the most important since wind power fluctuations at this horizon particularly affect the system balance, see \cite{Akhmatov2007} for instance. Emphasis here is on offshore wind power forecasting, since those short-term fluctuations in power generation are most significant offshore. Even though most efforts in wind power forecasting are placed on lead times from hours to days, many are investing in alternative approaches to improve the accuracy of very short-term forecasts, for instance leveraging detailed turbine-level data \cite{Gilbert2019}. Those very short-term lead times are not only crucial but also those it is the most difficult to improve the forecasts for, especially compared to the simple but very effective persistence benchmark. Forecasts characterize and reduce but do not eliminate uncertainty. Thus forecasts should be probabilistic in nature taking the form of probability distributions, following the argument of \cite{Dawid1984} among others.

Wind power generation is a stochastic process which is double-bounded by nature, both by zero when there is no production at all, and by the nominal power for high-enough wind speeds. For short-term forecasting, statistical methods have proved to be more skilled and accurate. However, those methods often rely on a Gaussian assumption -- which cannot be appropriate for a double-bounded variable. In \cite{Pinson2012}, it is proposed to move from the classical Gaussian assumption to a framework where the wind power variable follows a generalized logit-normal distribution. In this framework though, not all the parameters of the distribution are estimated and tracked, the shape parameter being selected upon cross-validation.

Consequently here, we propose to revisit this work and to estimate all the parameters of the generalized logit-normal distributions within a maximum likelihood framework. Such a framework is particularly suitable to obtain skilled probabilistic forecasts. In addition, emphasis is placed on describing both batch and recursive estimation approaches, in order to go towards an online learning approach as a basis for probabilistic forecasting. For a nice introduction to online learning, the reader is referred to \cite{Orabona2020}. Online learning (with exponential forgetting) makes it possible to accommodate the non-stationarity of wind power generation time-series. The models and estimation framework are first presented in Section \ref{sec:method}, and the resulting algorithms in Section \ref{sec:algos}. They are then applied to 10-min-ahead point and probabilistic forecasting at the Anholt offshore wind farm in Section \ref{sec:application}. Finally some concluding remarks and prospects are given in Section \ref{sec:conclusion}.

\section{Model and Estimation Framework}
\label{sec:method}
\subsection{Generalized Logit-Normal Distribution and its Parameters}
For an original random variable $X \in (0, 1)$, the generalized logit transform $Y$ is given by 
\begin{equation}
	Y=\gamma(X;\nu)=\ln\left(\frac{X^\nu}{1-X^\nu} \right), \quad \nu>0, 
\end{equation}
where $\nu$ is the shape parameter. When $Y$ follows a Gaussian distribution $\mathcal{N}(\mu,\sigma^2)$, the original variable $X$ follows a generalized logit-normal distribution $L_{\nu}(\mu,\sigma^2)$, see \cite{Pinson2012}. The probability density function is given by
\begin{equation}
\label{eq:f}
    f(x)=\frac{1}{\sqrt{2\pi\sigma^2}}\frac{\nu}{x(1-x^\nu)}\exp{\left[-\frac{1}{2}\left(\frac{\gamma(x;\nu)-\mu}{\sigma}\right)^2\right]}.
\end{equation}

Let $X$ the wind power random variable. We want $\nu$ such as the transform variable $Y$ is as close as possible to a Gaussian variable which then can be forecast in a Gaussian framework. As we have access to some realizations $(x_t)$ of $X$ and to the analytical expression of its density, we can then maximize the probability of observing the data $(x_t)$ depending on $\nu$, $\mu$ and $\sigma^2$, that is estimate all the parameters of the distribution \eqref{eq:f} using maximum likelihood inference. 

In the case of wind power generation, the observations $(x_t)$ are strongly correlated. We thus assume that  $Y_t|Y_{t-1},...,Y_{t-p} \sim \mathcal{N}(\mu_t,\sigma^2)$ where $\mu_t = \sum_{k=1}^p \phi_k Y_{t-k}$, that is the distribution of $X_t|X_{t-1},...,X_{t-p}$ is a generalized logit-normal distribution of density
\begin{equation}
\label{eq:f_conditional}
    \frac{1}{\sqrt{2\pi\sigma^2}}\frac{\nu}{x_t(1-x_t^\nu)}\exp{\left[-\frac{1}{2}\left(\frac{y_t-\sum_{k=1}^p\phi_k y_{t-k}}{\sigma}\right)^2\right]},
\end{equation}
where $y_t=\gamma(x_t;\nu)$. While the density in (\ref{eq:f_conditional}) is defined only for $x \in (0,1)$, the wind power generation can take values 0 and 1. We thus choose to look at the observations $x_t \in [0,1]$ as a coarsened version of $X$, see \cite{Lesaffre2007}. This coarsened data framework has been formalized by \cite{Heitjan1991} and \cite{Heitjan1993}.

\subsection{Maximum Likelihood Inference} 
Let $\Phi=(\phi_1,...,\phi_p)^\top \in \mathbb{R}^p$. The maximum likelihood inference is based on the likelihood function, given by 
\begin{equation}
\label{eq:L}
    L(\nu, \Phi, \sigma^2|\textbf{x})=\prod_{t=1}^{N}f(x_t|x_{t-1},...,x_{t-p}, \nu,\Phi,\sigma^2),
\end{equation}
which is the probability of the observed data under the model $f$, assuming the realizations of $X_t|X_{t-1},...,X_{t-p}$ are independent and identically distributed. We think of $L(\nu,\Phi,\sigma^2|\textbf{x})$ as a function of $\nu$, $\Phi$ and $\sigma^2$, the data $(x_t)$ being fixed.  The method of maximum likelihood chooses the values $(\nu,\Phi,\sigma^2)=(\hat{\nu},\hat{\Phi},\hat{\sigma}^2)$ to maximize $L(\nu, \Phi, \sigma^2|\textbf{x})$. The logarithm of $L$ being easier to maximize, especially when exponential distributions are involved, it is used instead of the likekihood. For model $f$ the negative log-likelihood function is
\begin{equation}
\label{eq:negl}
\begin{split}
    \Tilde{l}(\nu, \Phi, \sigma^2|\textbf{x}) & = \frac{N-p}{2}\ln(\sigma^2)-(N-p)\ln(\nu)
\\
    & +\sum_{t=p+1}^{N}\ln(1-x_t^\nu) 
\\    
    & +\frac{1}{2\sigma^2}(\textbf{y}-\textbf{Y}\Phi)^\top(\textbf{y}-\textbf{Y}\Phi) + C,
\end{split}
\end{equation}
where $\textbf{y}=(y_{p+1}, ..., y_N)^\top \in \mathbb{R}^{N-p}$, $\textbf{Y}$ is a matrix with columns $B\textbf{y}, B^2\textbf{y}, ..., B^p\textbf{y} \in \mathbb{R}^{(N-p) \times p}$, $B$ being the backshift operator, $C$ is a constant which does not depend on $\nu$, $\Phi$ or $\sigma^2$.
Computing the first derivatives of \eqref{eq:negl} w.r.t. the parameters of the distribution we can retrieve stationary points. It is worth noting that those points are minimizers only if the negative log-likelihood is convex. Taking the derivative of \eqref{eq:negl} w.r.t. $\Phi$, resp. $\sigma^2$, and setting it equal to zero, leads to the usual maximum likelihood estimators
\begin{equation}
\label{eq:Phi}
    \Hat{\Phi}=(\textbf{Y}^\top \textbf{Y})^{-1}\textbf{Y}^\top \textbf{y}, \quad \quad \Hat{\sigma}^2=\frac{(\textbf{y}-\textbf{Y}\Hat{\Phi})^\top (\textbf{y}-\textbf{Y}\Hat{\Phi})}{N-p}.
\end{equation}
Taking the derivative of \eqref{eq:negl} w.r.t. $\nu$, we thus need to solve
\begin{equation}
\label{eq:nu}
    -\frac{N-p}{\nu}-\sum_{t=p+1}^{N} \frac{\ln(x_t) {x_t}^\nu}{1-{x_t}^\nu} + \frac{(\textbf{u}-\textbf{U}\Phi)^\top(\textbf{y}-\textbf{Y}\Phi)}{\sigma^2} = 0,
\end{equation}
where $\textbf{u}=\frac{\partial \textbf{y}}{\partial \nu}$ with $u_t=\ln(x_t)(1+\frac{{x_t}^\nu}{1-{x_t}^\nu})$, $\textbf{U}=\frac{\partial\textbf{Y}}{\partial\nu}$ with columns $B\textbf{u}, B^2\textbf{u},...,B^p\textbf{u}$. Unlike $\hat{\Phi}$ and $\hat{\sigma}^2$, $\hat{\nu}$ does not have a closed-form solution and a descent algorithm is then to be used to solve \eqref{eq:nu}.

\section{Batch and Recursive Algorithms}
\label{sec:algos}
\subsection{Batch Algorithm} We use both the closed-form solutions in \eqref{eq:Phi} for $\Hat{\Phi}$ and $\Hat{\sigma}^2$, and a Newton-Raphson algorithm to solve \eqref{eq:nu} in order to estimate the shape parameter $\nu$. The computation of the Newton-Raphson step requires the second derivative of \eqref{eq:negl} w.r.t. $\nu$, i.e.
\begin{equation}
\label{eq:nu2}
\begin{split}
    \frac{\partial^2\Tilde{l}}{\partial\nu^2 } & = \frac{N-p}{\nu^2 }- \sum_{t=p+1}^{N}\ln(x_t)^2\frac{x_t^\nu}{(1-x_t^\nu)^2} 
\\
    & +\frac{(\textbf{v}-\textbf{V}\Phi)^\top(\textbf{y}-\textbf{Y}\Phi)}{\sigma^2} + \frac{\parallel \textbf{u}-\textbf{U}\Phi \parallel_2^2}{\sigma^2},
\end{split}
\end{equation}
where $\textbf{v}=\frac{\partial \textbf{u}}{\partial \nu}$ with $v_t=  u_t \ln(x_t) \frac{{x_t}^\nu}{1-{x_t}^\nu}$, $\textbf{V}=\frac{\partial\textbf{U}}{\partial \nu}$ with columns $B\textbf{v}, B^2\textbf{v},...,B^p\textbf{v}$.

The full algorithm is described in Algorithm \ref{algo:diagalgo} and has showed very fast convergence on numerous simulations of samples distributed according to the generalized logit-normal distribution with different values of $\Phi$, $\sigma^2$ and $\nu$.

\begin{algorithm}
\caption{Batch MLE with diagonalization}
\begin{algorithmic} 
\label{algo:diagalgo}
\STATE Set $i \leftarrow 1$ and let $\nu_1=1$, $\epsilon=0.001$.
\REPEAT 
\STATE{1. \textit{Update.} $\Phi_i=(\textbf{Y}^\top \textbf{Y})^{-1}\textbf{Y}^\top \textbf{y}$; $\sigma^2_i = \frac{(\textbf{y}-\textbf{Y}\Phi_i)^\top (\textbf{y}-\textbf{Y}\Phi_i)}{N-p}$.}
\STATE{2. \textit{Compute the Newton step and decrement for} $\nu$.
\newline $\Delta \nu_{nt}=-\frac{\nabla_\nu \Tilde{l}}{\nabla_\nu^2 \Tilde{l}}$; $\lambda^2=\frac{(\nabla_\nu \Tilde{l})^2}{\nabla_\nu^2 \Tilde{l}}$.}
\STATE{3. \textit{Stopping criterion.} \textbf{quit} if $\lambda^2/2 \leq \epsilon$.}
\STATE{4. \textit{Line search.} Choose step size $t$ by backtracking line search.}
\STATE{5. \textit{Update.} $\nu_{i+1}=\nu_{i}+t\Delta\nu_{nt}$.}
\UNTIL{termination test satisfied.}
\end{algorithmic}
\end{algorithm}

\subsection{Recursive Algorithm}
The batch algorithm is well suited if the data are known to be stationary to second order, that is assuming the parameters of the distribution \eqref{eq:f} do not change over the curse of time. But if, as we suspect in the case of wind power data, the time series is not stationary and the parameters are not constant, then the batch algorithm is not appropriate and alternative solutions are required. 
Recursive estimation allows for such a parametric time-variability and provides information not only on the existence of non-stationarity but also on the possible nature of the parametric variations (see e.g. \cite{Young1984}). 

As the inference relies on the likelihood function, it is straightforward to derive a recursive algorithm which only requires the first derivatives of \eqref{eq:negl} w.r.t. to the parameters. Let introduce $\Hat{\Theta}_t = (\Hat{\Phi}_t, \Hat{\sigma}^2_t, \Hat{\nu}_t)$ the estimate of the parameters at time $t$. The recursive estimation procedure relies on a Newton-Raphson step for obtaining the estimate $\Hat{\Theta}_t$ as a function of the previous estimate $\Hat{\Theta}_{t-1}$, see e.g. \cite{Madsen2007} and \cite{Madsen2012}. Let introduce the time-dependent negative log-likelihood objective function to be minimized at time $t$
\begin{equation}
\label{eq:recursiveobj}
    S_t(\Theta) = -\frac{1}{n_\alpha} \sum_{j=p+1}^{t} \alpha^{t-j}\ln(f_j(\Theta)),
\end{equation}
where $f_j(\Theta) = f(x_j | x_{j-1}, ... , x_{j-p}; \Theta)$, $\alpha$ is a forgetting factor, $\alpha \in (0,1)$, allowing for exponential forgetting of past observations, $n_\alpha = \frac{1}{1-\alpha}$ is the effective number of observations used for normalizing the weighted negative log-likelihood function. 
Applying one Newton-Raphson step we have
\begin{equation}
\label{eq:NRstep}
    \Hat{\Theta}_t = \Hat{\Theta}_{t-1} - \frac{\nabla_\Theta S_t(\Hat{\Theta}_{t-1})}{\nabla^2_\Theta S_t(\Hat{\Theta}_{t-1})}.
\end{equation}

As
\begin{equation}
\label{eq:gradient1}
\begin{split}
     \nabla_\Theta S_t(\Hat{\Theta}_{t-1}) & = \alpha \nabla_\Theta S_{t-1}(\Hat{\Theta}_{t-1})
\\
     &- (1-\alpha) \nabla_\Theta \ln(f_t(\Hat{\Theta}_{t-1})),
\end{split}
\end{equation}
assuming that $\Hat{\Theta}_{t-1}$ minimizes $S_{t-1}(\Theta)$, we get 
\begin{equation}
\label{eq:gradient2}
    \nabla_\Theta S_t(\Hat{\Theta}_{t-1}) = -(1-\alpha) \nabla_\Theta \ln(f_t(\Hat{\Theta}_{t-1})).
\end{equation}
From \eqref{eq:gradient1} we also get 
\begin{equation}
\label{eq:hessian1}
\begin{split}
     \nabla^2_\Theta S_t(\Hat{\Theta}_{t-1}) & = \alpha \nabla^2_\Theta S_{t-1}(\Hat{\Theta}_{t-1})
\\
     &- (1-\alpha) \nabla^2_\Theta \ln(f_t(\Hat{\Theta}_{t-1})).
\end{split}
\end{equation}

As
\begin{equation}
\label{eq:hessian2}
\begin{split}
    \nabla^2_\Theta \ln(f_t(\Hat{\Theta}_{t-1})) &= \frac{\nabla^2_\Theta f_t({\Hat{\Theta}_{t-1}})}{f_t({\Hat{\Theta}_{t-1}})} 
    \\
    &- \frac{\nabla_\Theta f_t(\Hat{\Theta}_{t-1})(\nabla_\Theta f_t(\Hat{\Theta}_{t-1}))^T}{f_t({\Hat{\Theta}_{t-1}})^2},
\end{split}
\end{equation}
assuming $f_t$ is (almost) linear in $\Theta$ in the neighborhood of $\Hat{\Theta}_{t-1}$, the first term in \eqref{eq:hessian2} vanishes and we obtain the following approximation
\begin{equation}
\label{eq:hessian3}
    \nabla^2_\Theta \ln(f_t(\Hat{\Theta}_{t-1})) = -\textbf{h}^{}_t \textbf{h}^\top_t,
\end{equation}
where $\textbf{h}_t = \frac{\nabla_\Theta f_t(\Hat{\Theta}_{t-1})}{f_t({\Hat{\Theta}_{t-1}})} = \nabla_\Theta \ln(f_t(\Hat{\Theta}_{t-1}))$.

\

Let $\Hat{\textbf{R}}_t = \nabla^2_\Theta S_t(\Hat{\Theta}_t)$ and assume that the objective criterion $S$ is smooth in the vicinity of $\Hat{\Theta}_{t}$, and the adaptation step small enough so that
\begin{equation}
\label{eq:assumption}
    \Hat{\textbf{R}}_t = \nabla^2_\Theta S_t(\Hat{\Theta}_t) \simeq \nabla^2_\Theta S_t(\Hat{\Theta}_{t-1}).
\end{equation}
This is a classic assumption for deriving recursive estimation methods for stochastic systems (see \cite{Ljung1983}). The two-step recursive scheme at time $t$ is then 
\begin{align}
    \Hat{\textbf{R}}_t &= \alpha \Hat{\textbf{R}}_{t-1} + (1-\alpha) \textbf{h}^{}_t \textbf{h}^\top_t, \label{eq:recursive1} \\
    \Hat{\Theta}_t &= \Hat{\Theta}_{t-1} + (1-\alpha) \Hat{\textbf{R}}_t^{-1} \textbf{h}_t. \label{eq:recursive2}
\end{align}
Equation \eqref{eq:recursive1} derives from \eqref{eq:hessian1} and \eqref{eq:hessian3}. Equation \eqref{eq:recursive2} derives from \eqref{eq:NRstep}, \eqref{eq:gradient2} and \eqref{eq:assumption}. The final algorithm is available in Algorithm \ref{algo:rMLE}.

\begin{algorithm}
\caption{Recursive MLE}
\begin{algorithmic} 
\label{algo:rMLE}
\STATE Let $\Phi_0=\textbf{0}, \sigma_0^2=1, \nu_0=1, \textbf{h}_0=\textbf{0}, R_0=0_{(p+2,p+2)}$. 
\REPEAT 
\STATE{1. \textit{Update.} $\Hat{\textbf{R}}_i = \alpha \Hat{\textbf{R}}_{i-1} + (1-\alpha) \textbf{h}^{}_i \textbf{h}^\top_i$.}
\STATE{2. \textit{Update.} $\Hat{\Theta}_i = \Hat{\Theta}_{i-1} + (1-\alpha) \Hat{\textbf{R}}_i^{-1} \textbf{h}_i$ if $i>100+p$.}
\UNTIL{$t$ the forecasting time.}
\end{algorithmic}
\end{algorithm}

\section{Very-short-term Wind Power Forecasting Application}
\label{sec:application}
We apply the proposed models to a real dataset consisting of wind power generation from a large wind farm, Anholt in Denmark, from July 1, 2013 to August 31, 2014. Emphasis is placed on the maximum likelihood framework and its online learning derivation. For a comparison of the generalized logit-normal distribution to other distributions (e.g., Beta) for the purpose of wind power forecasting, see \cite{Pinson2012}.

\subsection{Data Description}
Active power is available for 110 wind turbines at a temporal resolution of every 10 minute. The time series are scaled individually according to the nominal power of the wind turbines. The average generation over the wind farm is then computed depending on the number of wind turbines being available at each time step, in order to handle missing values. The resulting random variable is then $X_t \in [0,1]$, the average active power generated in the wind farm at time $t$. 

We are interested in forecasting $X_{t+1}$ (point forecasting) and its distribution (probabilistic forecasting) knowing the realization of $X_t$; the lead time is therefore 10-minute-ahead. We split our data into two datasets: 
\begin{itemize}
    \item a training/cross-validation dataset from July 1, 2013 to March 31, 2014, resulting in 39,450 observations,
    \item a test dataset from April 1 to August 31, 2014, resulting in 22,029 observations.
\end{itemize}
The training set is used to fit all models, the cross-validation set to select hyper-parameters if needed and the test set to compare the proposed methodology to the benchmarks. It is worth noting the training set is long enough for the Algorithm \ref{algo:rMLE} to be recursive yet on the training period, after a short warm-up of 100 iterations.

\subsection{Point Forecasting} 
\label{sec:point}
In order to evaluate and compare the performance of the proposed methods for point forecasting we use the Root Mean Square Error (RMSE). When a model requires hyper-parameters to be selected before estimating the parameters, we use the following procedure:
\begin{itemize}
    \item The candidate models are fitted over a grid of hyper-parameters' values from July 1 to October 31, 2013;
    \item they are then retrained in a time-series cross-validation scheme, from November 1, 2013 to March 31, 2014, for which the size of the training window increases as we evolve through the validation set (consistent with a leave-one-out setup);
    \item the hyper-parameters leading to the smallest RMSE on the cross-validation set are selected;
    \item finally the final model is fitted over the whole training/cross-validation set and used for forecasting on the test set.
\end{itemize}

\paragraph{Benchmarks} We compare our methods to three benchmarks: the persistence, a normal auto-regressive (NAR) model and its recursive version. The persistence consists in taking $\hat{x}_{t+1} = x_t$. The normal AR model assumes $X_t|X_{t-1},...,X_{t-p} \sim \mathcal{N}(\mu_t,\sigma^2)$ where $\mu_t = \sum_{k=1}^p \phi_k X_{t-k}$. In this Gaussian setup the forecasts are unbounded and happen to be greater than 1 or lower than 0. Thus we need to truncate \textit{a posteriori} the out-of-range predictions so they lie in the interval $[0,1]$. We test AR models up to lag $p=5$ and observe that no significant improvement is provided beyond lag 2 for both batch and recursive approaches. We thus select $p = 2$. For the recursive AR model we also need to select the forgetting factor $\alpha$, which exponentially weights data in the past. In a similar way, it is selected such as $\alpha = 0.995$.

\begin{figure}[!ht]
\centerline{
\includegraphics[width=.95\columnwidth]{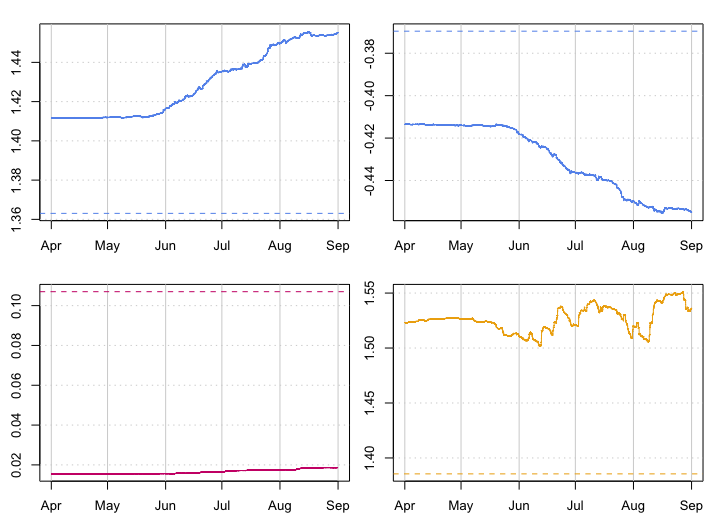}
}
\caption{Parameters of the generalized logit-normal distribution for $p=2$ and $\alpha=0.9994$: $\Hat{\Phi}$ (top), $\Hat{\sigma}^2$ (bottom left) and $\Hat{\nu}$ (bottom right).}
\vspace{-3mm}
\label{fig:ANH_point_parameters}
\end{figure}

\paragraph{Forecasting using generalized logit-normal  distributions} Let $\delta > 0$ such as each value being lower than $\delta$ (resp. greater than $1 - \delta$) is set to $\delta$ (resp. $1 - \delta$) and consider those "corrected" observations as the realizations of $X \in (0,1)$. In a symmetric way, forecasts being lower than $\delta$ (resp. greater than $1 - \delta$) will be set to 0 (resp. 1). $\delta$ is selected over cross-validation along with $p$. Algorithm \ref{algo:diagalgo} converges in 11 iterations towards the estimated values $\Hat{\nu}=1.39$, $\Hat{\Phi}=(1.363,-0.370)^\top$ and $\Hat{\sigma}^2=0.11$ for the selected combination of hyper-parameters  $\delta=0.005$ and $p=2$. For Algorithm \ref{algo:rMLE}, we choose $\delta=0.005$, $p=2$  and $\alpha=0.9994$ upon cross-validation. See in Fig. \ref{fig:ANH_point_parameters} the estimated parameters of the generalized logit-normal distributions over the test period. 

\paragraph{Results} The point forecasting performance over the test set of the benchmarks and the (GLNAR) proposed algorithms are available in Table \ref{tab1}. It is worth noting that the test set consists in 22,023 observations, which is a volume of data large enough to claim for significant results. The best point forecasts are obtained by the model using adaptive generalized logit-normal distributions. One can note that the model which uses a constant generalized logit-normal distribution gets poorer performance than the recursive AR model. Therefore the assumption that seems to matter the most here is the time-varying parameters assumption. Moreover, the estimated value of the scale parameter is significantly larger in the batch setup than in the recursive one, while the shape parameter is significantly lower. It may confirm that the recursive setup is more appropriate to the characteristics of the time series and thus allows for a better discrimination between the scale and the shape parameters of the distribution.
\begin{table}[!ht]
\vspace{-2mm}
\caption{10-minute-ahead RMSE over the test period, and respective improvements over persistence}
\begin{center}
\begin{tabular}{|c|c|c|}
\hline
\textbf{Model} & \textbf{RMSE}& \textbf{Imp. over persist.}\\
\hline
persistence &3.27\% &- \\
\hline
batch NAR &2.79\% &14.68\% \\
\hline
recursive NAR &2.72\% &16.82\% \\
\hline
batch GLNAR &2.74\% &16.21\% \\
\hline
recursive GLNAR &\textbf{2.70}\% &\textbf{17.43}\% \\
\hline
\multicolumn{3}{l}{*\textit{Best forecast bolded.}}
\end{tabular}
\label{tab1}
\end{center}
\end{table}
\vspace{-5mm}

\subsection{Probabilistic Forecasting} 
Let $F_t$ a predictive cumulative distribution function at time $t$. The Continuous Ranking Probabilistic Score (CRPS) is defined by
\begin{equation}
\label{eq:CRPS}
    \text{CRPS} = \frac{1}{T}\sum_{t=1}^T \text{crps}(F_t,x_t) = \int_{-\infty}^{\infty}\text{BS}(y)\text{d}y,
\end{equation}
where 
\begin{equation}
\label{eq:crps}
    \text{crps}(F_t,x_t) =  \int_{-\infty}^{\infty}\{F_t(y)-\textbf{1}(y \geq x_t)\}^2 \text{d}y,
\end{equation}
and BS is the Brier score
\begin{equation}
\label{eq:BS}
    \text{BS}(y) = \frac{1}{T}\sum_{t=1}^T \{F_t(y)-\textbf{1}(x_t \leq y)\}^2.
\end{equation}
See for example \cite{Brier1950} and \cite{Gneiting2007}. To evaluate the performance of the proposed models for probabilistic forecasting we use the CRPS instead of the RMSE, following the scheme described at the beginning of section \ref{sec:point}.
 
\paragraph{Benchmarks} We compare our method to four benchmarks: climatology, probabilistic persistence, and probabilistic versions of the batch and recursive AR models. Climatology consists in computing empirical quantiles on the training set. We test different grids and choose upon cross-validation to estimate the predictive cumulative distribution from the quantiles $\{0,0.01,...,0.99,1\}$. On the test set the quantiles are updated whenever a new observation is recorded. Probabilistic persistence consists in dressing the point persistence prediction with the most recent observed values of the persistence error. We choose the number of observed values upon cross-validation to be 20. For probabilistic AR forecasts, the least squares estimator of the variance of the residuals is used in both batch and recursive modes, and we assume those residuals to follow a Gaussian distribution $\mathcal{N}(0,\Hat{\sigma}^2)$. The forecast distribution of $x_t$ is then a Gaussian distribution $\mathcal{N}(\Hat{x}_t,\Hat{\sigma}^2)$ where $\Hat{x}_t$ is the point forecast from the AR model. The hyper-parameters $p$ and $\alpha$ for the recursive model are selected upon cross-validation with CRPS, which leads to $p=2$ as for point forecasting, but to a different $\alpha$ which is now equal to 0.983 instead of 0.995.

\begin{figure}[!ht]
\centerline{
\includegraphics[width=.95\columnwidth]{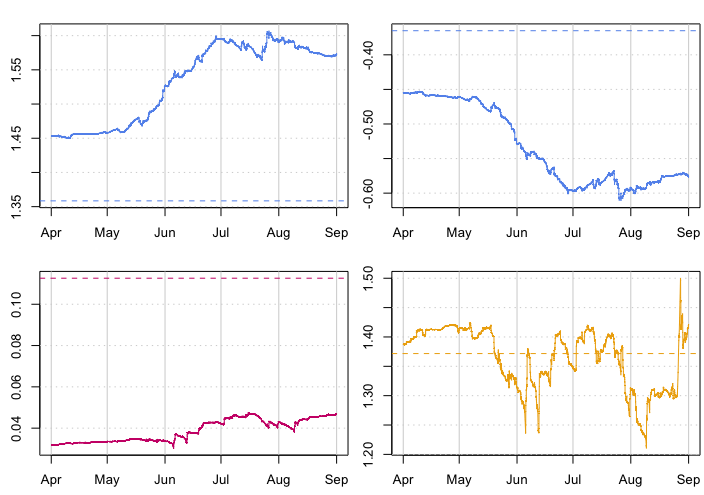}
}
\caption{Temporal evolution of the parameters of the generalized logit-normal distributions for $p=2$ and $\alpha=0.9994$: $\Hat{\Phi}$ (top), $\Hat{\sigma}^2$ (bottom left) and $\Hat{\nu}$ (bottom right).}
\label{fig:ANH_proba_parameters}
\vspace{-3mm}
\end{figure}

\paragraph{Forecasting using generalized logit-normal distributions} The lag $p$ selected upon cross-validation with CRPS remains equal to 2 in both batch and recursive algorithms, while $\delta$ and $\alpha$ change. For Algorithm \ref{algo:diagalgo}, now $\delta=0.006$ which leads to slightly different estimated parameters of the distribution: $\Hat{\nu}=1.37$ and $\Hat{\Phi}=(1.358,-0.365)^\top$, while the variance $\Hat{\sigma}^2=0.11$ remains the same. For Algorithm \ref{algo:rMLE}, now $\delta=0.004$ and $\alpha$ decreases from 0.9994 for point forecasting to 0.9986 for probabilistic forecasting. See in Fig. \ref{fig:ANH_proba_parameters} the estimated parameters of the generalized logit-normal distributions over the test period, which show higher time-variability because of the lower value of the forgetting factor. 

\begin{figure}[!ht]
\centerline{
\includegraphics[width=\columnwidth]{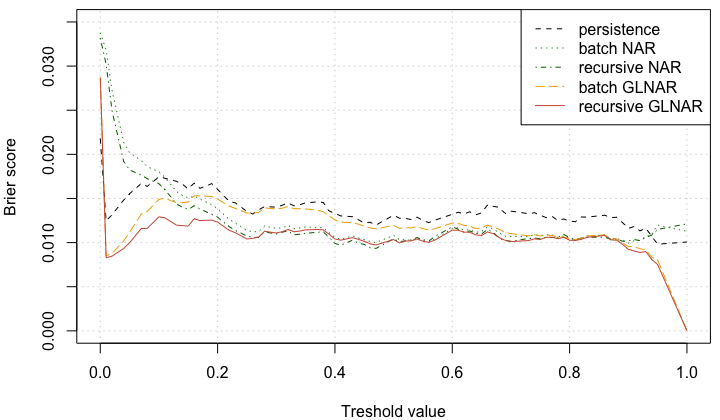}
}
\caption{Brier score computed over the test set for all methods but climatology, as a function of the chosen threshold.}
\label{fig:ANH_brier}
\end{figure}

\paragraph{Results} The CRPS computed over the test set for all the benchmarks and the proposed models are available in Table \ref{tab2}. The climatology's predictive cumulative distribution function $F_{t+1}$ remains unchanged whatever the value of $x_t$, which explains the very poor global performance of this method. The performance of the predictive cumulative distributions assuming a Gaussian setup and that of the approach using a constant generalized logit-normal distribution are close as for point forecasting. However, for probabilistic forecasting, the approach using adaptive generalized logit-normal distributions outperforms the other methods. The Brier scores are plotted in Fig. \ref{fig:ANH_brier}. As expected the methods using the generalized logit transformation perform better close to the bounds of the interval $[0,1]$. 
\begin{table}[!ht]
\vspace{-2mm}
\caption{10-minute-ahead CRPS over the test period, and respective improvements over climatology and persistence}
\begin{center}
\begin{tabular}{|c|c|c|c|}
\hline
\textbf{Model} & \textbf{CRPS} & \textbf{Imp. over clim.} & \textbf{Imp. over persist.}\\
\hline
climatology &22.04\% &- &-\\
\hline
prob. persistence &1.36\% &93.85\% &- \\
\hline
batch NAR &1.28\% &94.17\% &5.28\% \\
\hline
recursive NAR &1.23\% &94.34\% &9.40\% \\
\hline
batch GLNAR &1.21\% &94.52\% &10.90\%\\
\hline
recursive GLNAR &\textbf{1.06}\% &\textbf{95.17}\% &\textbf{21.57}\%\\
\hline
\multicolumn{4}{l}{*\textit{Best forecast bolded.}}
\end{tabular}
\label{tab2}
\end{center}
\end{table}
\vspace{-5mm}

\begin{figure}[!ht]
\centerline{
\includegraphics[width=\columnwidth]{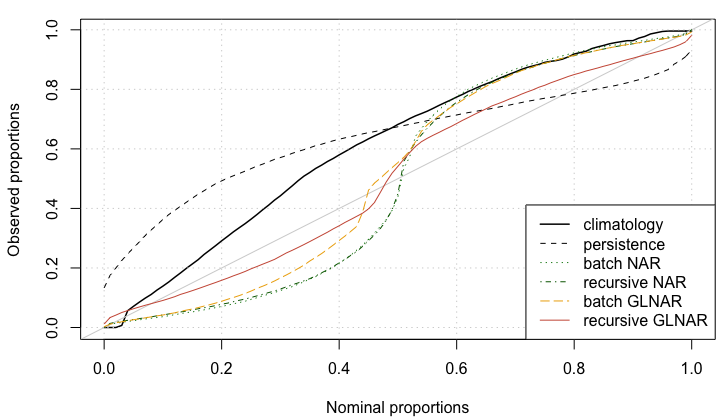}
}
\caption{Reliability diagram over the test set.}
\vspace{-3mm}
\label{fig:reliability}
\end{figure}

\begin{figure}[!ht]
\centerline{
\includegraphics[width=\columnwidth]{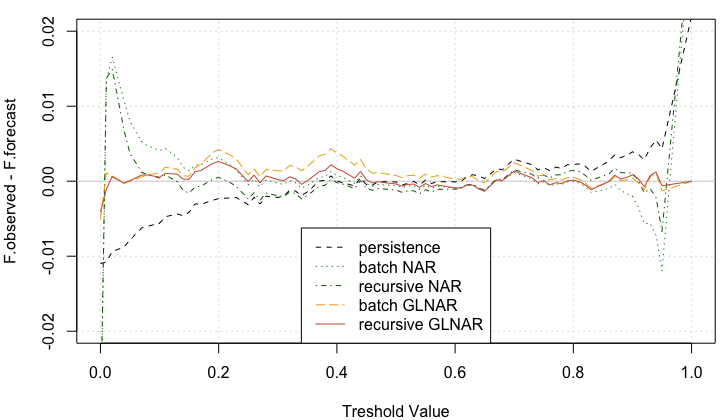}
}
\caption{Marginal calibration plot over the test set.}
\label{fig:marginal_calibration}
\vspace{-2mm}
\end{figure}

The CRPS and the Brier score give indications about the sharpness of the distributions. In order to check the calibration we show the results of two tools: the reliability diagram in Fig. \ref{fig:reliability} and a marginal calibration plot which is the difference between the average predictive $\bar{F}$ on the test set and the empirical cumulative distribution function in Fig. \ref{fig:marginal_calibration}. For the reliability diagram, the closer to the diagonal, the better the calibration, the empirical probabilities getting closer to the nominal ones. See \cite{Pinson2007} and \cite{Gneiting2007} for more details about those calibration tools. One can see that for both indicators the approach using adaptive generalized logit-normal distributions outperforms the other probabilistic forecasting methods. In Fig. \ref{fig:marginal_calibration} the climatology difference is not presented for being far bigger than zero.

\begin{figure}[!ht]
\centerline{
\includegraphics[width=\columnwidth]{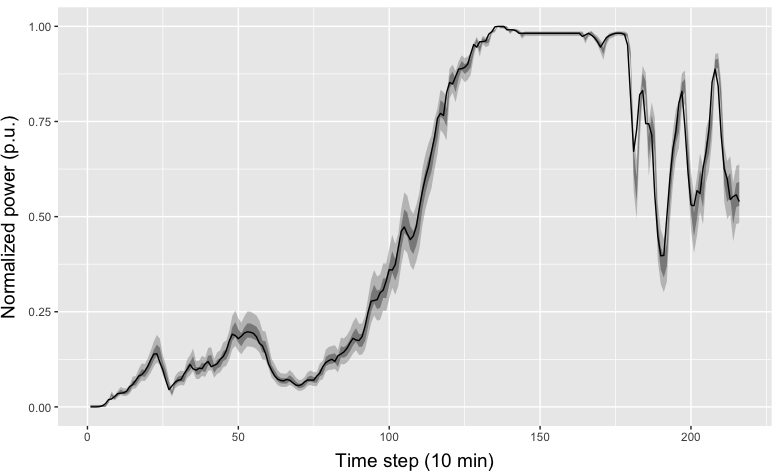}
}
\caption{Probabilistic forecasts from the recursive approach (Algorithm \ref{algo:rMLE}), based on prediction interval with nominal coverage rates of 95 and 75\%, along with the power measurements (solid black line).}
\label{fig:probapred}
\vspace{-3mm}
\end{figure}

Example probabilistic forecasts obtained from the adaptive generalized logit-normal approach over a 36 hour period of time are depicted in Fig. \ref{fig:probapred} by using prediction intervals with nominal coverage rates of 95 and 75\%.

\section{Conclusions}
\label{sec:conclusion} 
A generalized logit-normal distribution was considered for very short-term wind power forecasting, in order to adequately handle the double-bounded nature of the time series. All the parameters of the distribution were estimated from the data in a maximum likelihood framework, for both batch and online setups. The adaptive version of the distribution provides only a slight improvement in the accuracy of the point forecasts compared to approaches within a Gaussian framework, though it substantially outperforms the other benchmarks when focusing on probabilistic forecasting (intervals and full predictive densities). This confirms that such a choice of distribution may be most appropriate. While it achieves better calibration and sharpness, there is still room for improvement. In particular, we have emphasized the importance of the double-bounded nature of the process, but in practice the upper bound may also change in time. Indeed, wind power generation is not always bounded by the nominal capacity of the wind farm, e.g. in case of curtailment. It should then be taken into account within the modelling and forecasting framework, by additionally adaptively estimating this upper bound from data. 

Furthermore, the proposed framework could be applied for multi-step ahead forecasting, and makes it easy to assume other models for the conditional expectation of the transformed variable. In particular it is straightforward to add exogenous variables to the auto-regressive model, or to generalize it with a non-linear one. This may be a way to account for the individual productions of the wind turbines in order to improve the prediction of power generation for the whole wind farm. Finally, the $\delta$ hyper-parameter which handles the coarsened version of the distribution was selected upon cross-validation. It could instead enter a Bayesian or a likelihood inference as a parameter to be properly estimated.


\begin{thebibliography}{00}
\bibitem{Hong2020} T. Hong, P. Pinson, Y. Wang, R. Weron, D. Yang and H. Zareipour, "Energy forecasting: A review and outlook," IEEE Open Access Journal of Power and Energy, vol. 7, pp. 376-388, 2020.
\bibitem{Akhmatov2007} V. Akhmatov, ``Influence of wind direction on intense power fluctuations in large offshore wind farms in the North Sea,'' Wind Energ., vol. 31(1), pp. 59--64, 2007.
\bibitem{Gilbert2019} C. Gilbert, J. Browell and D McMillan, "Leveraging turbine-level data for improved probabilistic wind power forecasting,"
IEEE Trans. Sust. Energ., vol. 11, no. 3, pp. 1152--1160, 2019.
\bibitem{Dawid1984} A.P. Dawid, ``Statistical theory: the prequential approach'', J. R. Statist. Soc. A, vol. 157(2), pp. 278--292, 1984.
\bibitem{Pinson2012} P. Pinson, ``Very-short-term probabilistic forecasting of wind power with generalized logit-normal distributions,'' J. R. Statist. Soc. C, vol. 61(4), pp. 555--576, 2012.
\bibitem{Orabona2020} F. Orabona. A Modern Introduction to Online Learning. Lecture notes, Boston University, 2020.
\bibitem{Lesaffre2007} E. Lesaffre, D. Rizopoulous and R. Tsonaka, ``The logistic transform for bounded outcome scores,'' Biostatistics, vol. 8(1), pp. 72--85, 2007.
\bibitem{Heitjan1991} D. Heijtan and D. Rubin, ``Ignorability and coarse data,'' Ann. Stat., vol. 19(4), pp. 2244--2253, 1991.
\bibitem{Heitjan1993} D. Heijtan, ``Ignorability and coarse data: some biomedical examples,'' Biometrics, vol. 49(4), pp. 1099--1109, 1993.
\bibitem{Young1984} P. Young, Recursive estimation and time-series analysis: An introduction. Springer-Verlag, Berlin, Heidelberg, 1984.
\bibitem{Madsen2007} H. Madsen, Time Series Analysis. Chapman \& Hall, Boca Raton, 2007.
\bibitem{Madsen2012} P. Pinson and H. Madsen, ``Adaptive modelling and forecasting of offshore wind power fluctuations with Markov-switching autoregressive models,'' J. Forecast., vol. 31, pp. 281--313, 2012.
\bibitem{Ljung1983} L. Ljung and T. Söderström, Theory and Practice of Recursive Estimation, 1983.
\bibitem{Brier1950} G.W. Brier, ``Verification of forecasts expressed in terms of probability,'' Monthly Weather Review, vol. 78(1), pp. 1--3, 1950.
\bibitem{Gneiting2007} T. Gneiting, F. Balabdaoui and A.E. Raftery, ``Probabilistic forecasts, calibration and sharpness,'' J. R. Statist. Soc. B, vol. 69(2), pp. 243--268, 2007.
\bibitem{Pinson2007} P. Pinson, H.Aa. Nielsen, J.K. Møller and H. Madsen, ``Non-parametric probabilistic forecasts of wind power: Required properties and evaluation,'' Wind Energ., vol. 10(6), pp. 497--516, 2007.
\end{thebibliography}
\end{document}